\documentclass[prb,twocolumn,superscriptaddress,amsmath,amsfonts,nofootinbib]{revtex4-2}

\usepackage{graphicx}

\usepackage{color}

\newcommand{\ket}[1]{\left\vert#1\right\rangle}

\begin{document}

\author{Henrik Dreyer}
\affiliation{Max-Planck-Institute of Quantum Optics, Hans-Kopfermann-Stra\ss{e}~1, 85748 Garching, Germany}
\affiliation{Munich Center for Quantum Science and Technology, Schellingstra\ss{}e~4, 80799 M\"unchen, Germany}
\author{Laurens Vanderstraeten}
\affiliation{Department of Physics and Astronomy, Ghent University, Krijgslaan 281, S9, 9000 Gent, Belgium}
\author{Ji-Yao Chen}
\affiliation{Max-Planck-Institute of Quantum Optics, Hans-Kopfermann-Stra\ss{e}~1, 85748 Garching, Germany}
\affiliation{Munich Center for Quantum Science and Technology, Schellingstra\ss{}e~4, 80799 M\"unchen, Germany}
\author{Ruben Verresen}
\affiliation{Department of Physics, Harvard University, Cambridge MA 02138, USA}
\author{Norbert Schuch}
\affiliation{Max-Planck-Institute of Quantum Optics, Hans-Kopfermann-Stra\ss{e}~1, 85748 Garching, Germany}
\affiliation{Munich Center for Quantum Science and Technology, Schellingstra\ss{}e~4, 80799 M\"unchen, Germany}
\affiliation{University of Vienna, Faculty of Mathematics, Oskar-Morgenstern-Platz 1, 1090 Wien, Austria}
\affiliation{University of Vienna, Faculty of Physics, Boltzmanngasse 5, 1090 Wien, Austria}

\title{Robustness of critical $\mathbf{U(1)}$ spin liquids 
and emergent symmetries in tensor networks}

\begin{abstract}
We study the response of critical Resonating Valence Bond (RVB) spin
liquids to doping with longer-range singlets, and more generally of
$\mathrm{U}(1)$-symmetric tensor networks to non-symmetric
perturbations.
Using a field theory description, we find that in the RVB, doping constitutes a
relevant perturbation which immediately opens up a gap, contrary to
previous observations. Our analysis predicts a very
large correlation length even at significant doping, 
which we verify using high-accuracy numerical simulations. 
This emphasizes the need for careful
analysis, but also justifies the use of such states as a
variational ansatz for critical systems. 
Finally, we give an example of a
PEPS where non-symmetric perturbations do not open up a gap and the
$\mathrm{U}(1)$ symmetry re-emerges.
\end{abstract}

\maketitle

Projected Entangled Pair States (PEPS) form a powerful analytical and
numerical framework for describing strongly correlated quantum systems,
such as spin liquids or systems with topological
order~\cite{verstraete:2D-dmrg,verstraete:comp-power-of-peps,buerschaper:stringnet-peps,gu:stringnet-peps,bridgeman:interpretive-dance,cirac:tn-review-2009,schollwoeck:review-annphys}.
Their power stems from the local description where a tensor correlates
physical and entanglement degrees of freedom.  A key strength of PEPS is
the encoding of physical symmetries in symmetries of the tensor, which
allows to locally impose, probe, and control desired properties, and is
central to applications ranging from the classification of phases all the way
to efficient
algorithms~\cite{perez-garcia:inj-peps-syms,molnar:normal-peps-fundamentalthm,chen:1d-phases-rg,schuch:mps-phases,chen:spt-order-and-cohomology,chen:2d-spt-phases-peps-ghz,jiang:sym-peps-phases,schollwoeck:review-annphys,weichselbaum:nonabelian-syms-in-tn,mambrini:su2-peps-classes}.

However, not only physical symmetries are reflected in the tensor: PEPS
can exhibit symmetries acting purely on the entanglement degrees of
freedom, which are deeply
connected to both topological order and critical behavior, and closely
tied to a Gauss law of the underlying field theory.  In
particular, topological order in 2D is accompanied by entanglement
symmetries which act as representations of a discrete group (or some more 
general algebraic structure), such as $\mathbb Z_2$ for the Toric Code
model~\cite{schuch:peps-sym,buerschaper:twisted-injectivity,sahinoglu:mpo-injectivity}.
Another such connection is between continuous symmetries, in
particular $\mathrm{U}(1)$, and criticality.  A key example where this
occurs are the dimer model and the spin-$\tfrac12$ RVB state on
bipartite lattices, which has been studied in depth since Anderson
proposed it as an ansatz wavefunction for the parent state of the high-Tc
cuprate superconductors~\cite{anderson:rvb-highTC,moessner:quantum-dimer-models}.

But how closely are these symmetries linked to the physics observed---are
they strictly necessary, or do they just happen to appear in the specific
PEPS representation used?  This is of central importance for the
construction of variational ansatzes, since it determines whether we need
to stabilize said symmetry to be able to capture certain physics, such as
topological order or criticality. In the case of topological order,
breaking the discrete entanglement symmetry induces doping with
quasiparticles which in 2D immediately destroys topological order beyond a
certain length scale, 
just as finite temperature~\cite{chen:topo-symmetry-conditions,balents:var-states-topological}
(but not in 3D~\cite{delcamp:3d-toriccode-peps,williamson:3d-toriccode-peps}),
and thus, hardwiring those symmetries is essential to obtain a
wavefunction with true topological order.

For continuous entanglement symmetries such as $\mathrm{U}(1)$ in critical systems, the
situation is less clear. For instance, in the RVB and dimer
model, doping with longer-range (LR) singlets will generally break the
$\mathrm{U}(1)$ symmetry. However, for PEPS models which realize such 
doping, evidence for an extended critical regime up to significant 
doping has been observed when studying them as variational ansatzes for 
frustrated Heisenberg models, such as the $J_1$-$J_2$-model on the square
lattice~\cite{wang:rvb-square-lattice,chen:topo-z2-u1}.  This
raises several questions: Can one refrain from stabilizing the
$\mathrm{U}(1)$ symmetry when aiming
for a critical wavefunction? Could stability of the critical phase point
to an \emph{emergent $\mathrm{U}(1)$ symmetry}, something not yet 
observed in PEPS, and different from what is seen for 2D PEPS with 
discrete symmetries and topological order?  Finally, can we
understand this in terms of the underlying field theory, just as the
breakdown of topological order under perturbation can be explained from
quasiparticle doping?

In this paper, we study critical spin liquids with a PEPS
representation with a $\mathrm{U}(1)$ entanglement symmetry,
and investigate their robustness under perturbations away
from the $\mathrm{U}(1)$ point, with the RVB state with
LR singlets as our guiding example. To this end, we employ an
effective field theory description, treating the transfer matrix as a
Luttinger liquid with parameter $K$.  This allows us to analyze
the perturbations away from the $\mathrm{U}(1)$ point as 
perturbations in the field theory.
Applying this
to the RVB and dimer PEPS doped with LR singlets reveals that
this is a relevant perturbation and thus should open up a gap immediately.
However, a scaling analysis reveals
that for the RVB state, the gap opens up extremely slowly, which explains
why this gap has not been observed in previous simulations. We support our
analysis by high-precision numerics, from which we can reliably extract 
correlation lengths on the order of $10^4$ sites, far beyond what had been
observed before. The results match well with the scaling analysis as
well as a more quantitative prediction based on the sine-Gordon model. 
Our findings are not limited to the RVB model, but
apply generally to PEPS with a $\mathrm{U}(1)$ entanglement symmetry
and $K>\tfrac12$.

We conclude by discussing ways to obtain models where such perturbations
do not open up a gap and thus give rise to an emergent $\mathrm{U}(1)$
symmetry. We provide an example of a PEPS wavefunction where  field theory
predicts such a robustness, and give numerical evidence that 
under $\mathrm{U}(1)$-breaking perturbations, the $\mathrm{U}(1)$
symmetry re-emerges. This constitutes the
very first observation of emergent symmetries in PEPS.

\begin{figure}[b]
\includegraphics{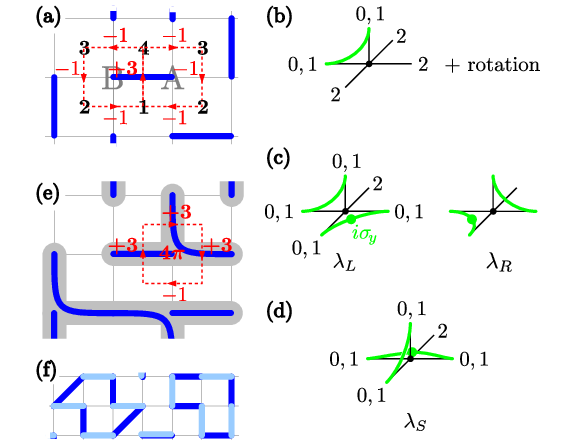}
\caption{\textbf{(a)} RVB and dimer model. Numbers in the plaquettes are
the height potential $h(\vec x)$ obtained from the $\mathrm{U}(1)$ Gauss
law (red). \textbf{(b)} PEPS tensor for the RVB state. The green line is
the identity on the $\{\ket{0},\ket{1}\}$ space, and all rotations are
summed.  \textbf{(c)} Tensors with ``teleportation bonds'' which give rise
to longer-range singlets; the two tensors are related by reflection.
\textbf{(d)} For $-\lambda_L=\lambda_R$,
the sum of the two
tensors in (c) is equivalent to the tensor shown with only straight
teleportation bonds, explaining the observed absence of diagonal
AA-singlets~\cite{wang:rvb-square-lattice}. \textbf{(e)}~Model with longer-range
singlets (c), and the ``dimer-solidomer'' model (gray). The teleportation
tensors carry a flux $2\pi m = 4\pi$. \textbf{(f)} Overlap of two
singlet configurations in the RVB, forming loops. Loops must contain an
even number of same-sublattice singlets. Since configurations with more
loops are favored, in expectation values same-sublattice singlets tend to
come in (ket-ket or ket-bra) pairs.  }
\label{fig:models}
\end{figure}

Let us start by introducing the RVB and dimer
model~\cite{anderson:rvb-highTC,moessner:quantum-dimer-models}. Throughout the paper,
it will serve as our guiding example, even though our key findings 
apply in generality. 
The dimer model on the square lattice is the equal
weight superposition $\ket\Psi=\sum\ket{D}$ of all coverings $\ket{D}$ of the
lattice with nearest neighbor dimers (Fig.~\ref{fig:models}a); the RVB
model is obtained by replacing the dimers by spin-$\tfrac12$ singlets $\ket{\sigma(D)}$,
oriented from the A to B sublattice.  The RVB and dimer model have 
natural PEPS representations,
Fig.~\ref{fig:models}b~\cite{verstraete:comp-power-of-peps,schuch:rvb-kagome}: The tensor
$P^i_{lurd}$ is constructed such that the \emph{physical index} $i=0,1$ is
identified with any one of the four \emph{virtual indices}
$l,r,u,d=0,1,2$, while the other three take the
value $2$, such that the tensor has the point group symmetry.
  By arranging these tensors on a
square grid and contracting adjacent virtual indices with a singlet
 inbetween, we obtain the RVB
state. Similarly, we can construct a PEPS for the dimer model by adding
another physical index at each site, which duplicates the information along
which direction the singlet is
placed~\cite{schuch:rvb-kagome}. We label these two settings by 
$\ket{\sigma_g(D)}$, 
$\ket{\Psi_g}=\sum \ket{\sigma_g(D)}$,
where $g=0$ corresponds to the RVB and $g=1$ to the dimer model.

The RVB and dimer model naturally possess a $\mathrm U(1)$ Gauss law:
given any region with an identical number of A and B sublattice sites,
consider all singlets which cross the boundary of the region, and count
how many of them cover an A sublattice vertex inside the region, vs.\
how many cover a B sublattice vertex. The difference of these two
numbers is always zero, since each singlet which sits fully inside the region
covers one A and one B vertex each, leaving an equal number of A and B
vertices to be paired up with outside vertices.  In the PEPS
representation, the Gauss' law can be seen as follows, illustrated in
Fig.~\ref{fig:models}a: Each tensor has exactly three virtual indices in
the state $\ket{2}$ (i.e., no singlet), while the forth is in the subspace
spanned by $\{\ket{0},\ket{1}\}$ (singlet); if we assign a weight $-1$ to
each $\ket{2}$ state and a weight $+3$ to states in the
$\{\ket{0},\ket{1}\}$ subspace, this adds up to zero and thus gives rise
to a $\mathrm{U}(1)$ entanglement symmetry of the tensor. 
This allows to derive a
potential, the \emph{height
representation}~\cite{fradkin2013field,alet:dimers-with-interactions,tang:rvb-qmc},
where we assign 
to every plaquette 
a height $h(\vec x)$ 
which changes by $+3$ when crossing a dimer
clockwise (counterclockwise) around an A (B) sublattice vertex, and by $-1$ otherwise.  
While for expectation values
$\langle\Psi_g\ket{\Psi_g}=\sum_{D,D'}\langle\sigma_g(D)\ket{\sigma_g(D')}$,
separate height fields $h$, $h'$ are associated to ket and bra, it has
been demonstrated in Ref.~\cite{damle:rvb-as-interacting-dimer} that they are
locked together in the long-wavelength limit and can thus be replaced by a
single field. Intuitively, this can be understood as follows (we refer the
interested reader to Ref.~\cite{damle:rvb-as-interacting-dimer}, where a
detailed treatment is given): For the dimer model, this is
obvious since $\langle D'\ket{D}=\delta_{D,D'}$ and thus
$h'=h+\mathrm{const.}$, and for the 
RVB model, configurations where $D$
and $D'$ differ more than locally give rise to longer and thus to less
loops in the overlap $\langle\sigma_g(D)\ket{\sigma_g(D')}$ 
(Fig.~\ref{fig:models}f) which are thus supressed (each loop gives a factor 
$2$ in $\langle\sigma_g(D)\ket{\sigma_g(D')}$); a more formal argument
follows
Ref.~\onlinecite{damle:rvb-as-interacting-dimer}, where
$\langle\Psi_g\ket{\Psi_g}$ is mapped to a dimer model with an irrelevant
interaction.

Since moreover changing $h(\vec x)\to h(\vec x)+4$ in a region leaves the pattern
locally invariant, this leads us to a compactified field (see
Refs.~\cite{fradkin2013field,alet:dimers-with-interactions,tang:rvb-qmc}
for a detailed discussion)
$\phi(\vec x)=\tfrac\pi2h(\vec x)\in[0;2\pi)$ governed by an effective
$(2+0)$D field theory
which captures the long-wavelength physics of the dimer and RVB
model~\cite{fradkin2013field,alet:dimers-with-interactions,tang:rvb-qmc},
\begin{equation}
\label{eq:action}
S_{\mathrm{free},K}= \frac{1}{8\pi K}\int\mathrm{d}\vec x\,(\vec\nabla\phi)^2\ ,
\end{equation}
where for the dimer model, $K=1$ is known analytically.

The RVB PEPS with tensor $P$ can be naturally generalized to include
LR singlets, by adding a tensor $Q$ 
consisting of terms (see Fig.~\ref{fig:models}c) which additionally entangle
two of the virtual indices through a singlet (a ``teleportation bond'')
around corners
with suitable weights $\lambda_L$ and $\lambda_R$,
and thus give rise to singlets between non-NN
sites~\cite{wang:rvb-square-lattice,chen:topo-z2-u1}.
Specifically, Ref.~\onlinecite{chen:topo-z2-u1} chooses
$\lambda_L=\lambda_R>0$, while Ref.~\onlinecite{wang:rvb-square-lattice} chooses
$-\lambda_L=\lambda_R>0$; we cover both with a single parameter
$\lambda$, where $\lambda_L=\lambda$ and
$\lambda_R=|\lambda|$, and additionally define
$\tilde\lambda=\sqrt{6}\lambda$ (the parameters used  in
Refs.~\onlinecite{wang:rvb-square-lattice} and \onlinecite{chen:topo-z2-u1},
respectively).\footnote{Note that
for $\lambda<0$, adding the two terms in Fig.~\ref{fig:models}c yields an
equivalent representation with only straight teleportation bonds
with weight $\lambda_S=\lambda_R=-\lambda_L$,
Fig.~\ref{fig:models}d. This explains why
Ref.~\onlinecite{wang:rvb-square-lattice} did not observe AA-sublattice singlets
between diagonally adjacent sites, and shows that the resulting LR-doped
RVB state is very special as it contains LR-singlets solely along the
lattice axes.
}
The perturbation $P\to P+\lambda Q$ breaks the 
$\mathrm{U}(1)$ entanglement symmetry of the tensors down to a $\mathbb Z_2$ symmetry, since the
number of virtual $2$'s now can be either $1$ or $3$; 
on the physical degrees of freedom, this is reflected in the fact that it
induces same-sublattice singlets, Fig.~\ref{fig:models}e.
To mimick this doping in the dimer model, we allow for trivalent objects
at vertices (shown gray in Fig.~\ref{fig:models}e)
in addition to dimers,
which breaks $\mathrm{U}(1)\to\mathbb Z_2$.
We
call this the dimer-solidomer model; it obeys a $\mathbb
Z_2$ Gauss' law and can thus be mapped to a $\mathbb Z_2$ loop or vertex model.

\begin{figure}
\includegraphics{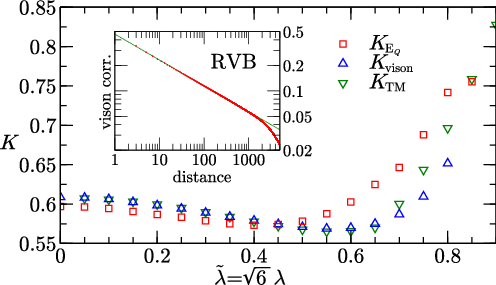}
\caption{
Luttinger parameter $K$ obtained via different methods:
$\mathbb E_Q$ and vison correlators ($K_{\mathbb E_Q}$,
$K_\mathrm{vison}$), and finite size extrapolation ($L=6,8$) of the
transfer matrix spectrum ($K_\mathrm{TM}$). The good agreement up to
$\lambda\approx 0.5$ is consistent with an underlying critical theory.
Inset: Decay of vison correlations for the RVB ($\lambda=0$), obtained
with iMPS bond dimension $\chi=2396$.  The correlations show perfect
algebraic decay up to a distance of above $1000$ sites.
}
\label{fig:K}
\end{figure}

For the above PEPS ansatz of doping the RVB model with LR singlets, strong
indications for critical behavior up to $0\le\tilde\lambda\le 0.85$ and for
$\tilde\lambda\approx -0.85$
 have been observed, as witnessed by an
algebraic decay of the correlation functions and extraction of a central
charge $c=1$ from a suitable
scaling~\cite{wang:rvb-square-lattice,chen:topo-z2-u1}.  This suggests
that the model for $\lambda\ne0$ should indeed be described by an
effective theory of the form \eqref{eq:action},
possibly with a different Luttinger parameter $K$.  
To further strengthen this point, we extract 
$K$ in different ways and check for consistency.  

On the one hand, we
determine $K$ from different two-point correlations. Each
operator contains primary fields $e^{i (e \phi + m\theta)}$ with charge
$e$ and flux $m$, with associated scaling dimension
\begin{equation}
\label{eq:scalingdim}
[e^{i(e\phi+m\theta)}] \equiv \Delta_{e,m} = Ke^2 + \frac{1}{4K}m^2 \ ,
\end{equation}
where we expect all fields allowed by symmetry considerations to appear.
($[O]$ denotes the scaling dimension of $O$, and 
$\theta$ is the dual field, $\partial_i\phi=
2K\varepsilon_{ij}\partial_j\theta$.)
We use two correlation functions: First, between solitons
(i.e.\ visons, living on plaquettes), which correspond to applying a $-1$ phase
for each dimer along a cut starting at plaquette $\vec
x_0$, or in the PEPS a string of $Z=\mathrm{diag}(-1,-1,1)$ 
placed on the bonds, where we compute the overlap with the vacuum;
note that this correlator is typically not accessible with other methods. 
Changing between
dimer and no dimer around $\vec x_0$ corresponds to changing $h(\vec x_0)$
by $\pm 4$, and thus, the resulting minus sign corresponds to an operator
$e^{\pm i\phi(\vec x_0)/2}$ in the field theory, and thus an electric charge
$|e|=\tfrac12$.  
Second, we consider changing the tensor at a given vertex to
one with a reduced $\mathbb Z_2$ symmetry---specifically, a solidomer in
the dimer model or a ket-bra pair of $Q$ tensors, which we call $\mathbb
E_Q$, for the RVB; 
since we change $-1 \to +3$ twice,
this corresponds to a magnetic flux $m=\tfrac{1}{2\pi}
\oint \vec\nabla\phi\cdot\mathrm{d}\vec r = \pm2$, Fig.~\ref{fig:models}e. 
(Since we place the same flux in both ket and bra, the effective
field theories of ket and bra vector in the long-wavelength limit can still
be described jointly.)
The correlation function between a pair of operators with scaling
dimension $\Delta$ decays as $\ell^{-2\Delta}$ with the separation $\ell$,
and thus as
$\ell^{-K/2}$ and $\ell^{-2/K}$, respectively, for the perturbations
considered, which allows us to extract
$K$ from numerical boundary iMPS simulations (extrapolating in the
finite correlation length induced by the iMPS bond dimension).

In addition, we extract $K$ from the transfer matrix $\mathbb T$,
that is, a ``slice'' of $\langle\Psi_g\ket{\Psi_g}$, on cylinders
of circumference $L$. In the IR, we expect that $\mathbb T \sim e^{-H}$,
where the field theory of $H$ is the Wick rotated version of \eqref{eq:action};  
for the dimer model, an exact mapping to free fermions and thus a 
Luttinger liquid with $K=1$ is indeed
known~\cite{lieb:dimer-freeferm,suzuki:dimer-top-to-ferm}.
 It is well
known that the energy spectrum of this $(1+1)$D theory is given by
the primaries $E_{e,m}=E_0(L)+\alpha\,\Delta_{e,m}/L$ for integer $e$, $m$ and
their descendants~\cite{difrancesco:CFTbook}.  For $\tfrac12\le K\le1$, the two dominant subleading
eigenvalues of $\mathbb T$ correspond to $E_{1,0}$ and $E_{0,1}$, 
such that we can extract $K$ as 
$4K^2=(E_{1,0}-E_{0,0})/(E_{0,1}-E_{0,0})$.

The results obtained from all three methods are shown in Fig.~\ref{fig:K} for the
RVB doped with LR singlets; in particular, this also yields an
estimate $K=0.609(10)$ for the nearest neighbor RVB model,
in agreement with earlier estimates from 
Monte Carlo
simulations~\cite{tang:rvb-qmc,albuquerque:rvb-square-correlations}.
 The
good agreement between the results obtained with different
methods strengthens the case for a description of the LR-doped model in
terms of the effective field theory
\eqref{eq:action}, or equivalently a Luttinger liquid.

The applicability of the effective CFT description \eqref{eq:action} suggests that we
should be able to use it to assess stability of the critical phase under
doping with LR singlets, $P\to P + \lambda Q$.  Such perturbations can
either be relevant -- they open up a gap, and have $\Delta<2$ -- or
irrelevant -- they disappear under RG, leaving the system critical, and
have $\Delta>2$.  If we could show that the perturbation $P+\lambda
Q$ which includes LR singlets was irrelevant, this would provide a
compelling explanation of 
the apparent criticality.

To test this hypothesis, we therefore need to determine the scaling dimension
of the perturbation $P+\lambda Q$. It is given by the 
subleading term when expanding $\langle\Psi_g\ket{\Psi_g}=\sum
\langle\sigma_g(D)\ket{\sigma_g(D')}$ in orders of $\lambda$.  The effect of a $Q$ tensor on the
A (B) sublattice is to induce a BB (AA) sublattice singlet.  Terms
$\langle\sigma_g(D)\ket{\sigma_g(D')}$ with a single $Q$ tensor 
vanish: Each term in the sum is an overlap of singlets, which form closed
loops (Fig.~\ref{fig:models}f); in this overlap, a same-sublattice singlet
must be always accompanied by another such singlet.  The leading
non-zero term is thus second order and consists of pairs of $Q$ tensors.
These pairs can appear in two ways: Either an AA and a BB pair in the same
(say, ket) layer, or an AA (or BB) pair both in the ket and the bra layer.
In both cases, those singlets will be bound together: Otherwise, a long
loop appears in the overlap which supresses it.
This can also be understood from symmetry considerations: As we
have seen, a single $Q$ tensor on the A (B) sublattice has flux
$m=\pm2$, and since in the long-wavelength limit, the magnetic potential
$\phi$ for ket and bra are locked, magnetic fluxes (and thus
same-sublattice singlets) must come in pairs with equal ket and bra flux.
Moreover, this shows that 
ket-bra pairs of $Q$ have flux $m=2$ while ket-ket pairs have flux $m=0$.
In addition, these pairs can also exhibit a non-trivial charge. However,
those charges are restricted by lattice symmetries:
The smallest non-trivial charge consistent with
$C_{4v}$ is $e=4$~\cite{alet:dimers-with-interactions}, which by itself is
an irrelevant perturbation for $K>1/8$.

It follows that the only potentially relevant perturbation which remains are bound ket-bra pairs
of $Q$ operators on the same sublattice.  Those have flux $m=2$ a
thus a scaling dimension $\Delta=1/K$.  One might wonder whether there
could be cancellation effects (either with pairs nearby, or between 
$Q$'s located at different distances), 
but we have found that summing those correlations 
converges quickly while not changing our findings, and re-grouping does
not give rise to cancellations; this is consistent with the fact that
such perturbations are allowed by symmetry.  We thus find that
in second order,
doping with LR-singlets can be understood as adding a perturbation with scaling
dimension
$\Delta=1/K$.  However, for the observed values of $K>\tfrac12$, this
implies that $\Delta<2$---that is, from a field
theory perspective, the perturbation is in fact relevant and \emph{should}
open up a gap!

Thus, we find---rather surprisingly---that the effective field theory,
which on the one hand seems to very well describe the system at hand, does
not explain the existence of an extended critical regime
when doping with longer-range singlets,
  but rather predicts the opening of a gap. So why has it not
been observed, and why did our initial tests further support 
critical behavior?  To understand this, let us carry out a scaling
analysis.  In leading order, the perturbed model is of the form
\begin{equation}
\label{eq:sinegordon}
S(\lambda) = \frac{K(\lambda)}{2\pi}\int \mathrm{d}^2x\, (\nabla\theta)^2
 + \omega \lambda^2 \int\mathrm{d}^2 x\, \cos(2\theta)\ ,
\end{equation}
where $\omega$ is a yet unknown parameter which relates the second-order
perturbation in $Q$ to $\cos(2\theta)$,
and where we have rewritten the free part in terms of the dual
field.  Note that $K(\lambda)$ has been renormalized due to
marginal terms in the perturbation, as we have observed in
Fig.~\ref{fig:K}.

Since $[e^{\pm i\,2\theta}]=1/K$, $[\int\mathrm{d}^2x\,\cos(2\theta)]
=-2+1/K$,
and thus for the action to be scale-invariant, $\sqrt{\omega}\lambda$ must scale as
$[\sqrt{\omega}\lambda]=1-1/2K$. On the other hand, $[\xi]=-1$,
 and thus
\begin{equation}
\label{eq:xi-scaling}
\xi \propto (\sqrt{\omega}\lambda)^{-u}\mbox{\ with\ } u=\left(1-\tfrac{1}{2K}\right)^{-1} 
= \frac{1}{2(K-\tfrac12)}+1\,.
\end{equation}
For the dimer model ($K=1$), we find that $u=2$, while for the RVB state
with $K\approx 0.6$, $u\approx 6$, that is, the correlation
length diverges rapidly in the perturbation 
as long as $\omega$ does not change too much, and will
thus exceed the previously observed bounds on $\xi$ of a few hundred sites
already for moderately large $\lambda$.
 Note that this also implies that
on length scales sufficiently below $\xi$ and above the lattice spacing,
the system is well described by the free theory. This justifies using
the $K(\lambda)$ obtained earlier from fitting the critical correlations 
at distances sufficiently below $\xi$, Fig.~\ref{fig:K}
(i.e., the UV of the sine-Gordon model)
to fit $K(\lambda)$ in Eq.~\eqref{eq:sinegordon}, cf.\
Ref.~\cite{lukyanov:sine-gordon}.

\begin{figure}
\includegraphics{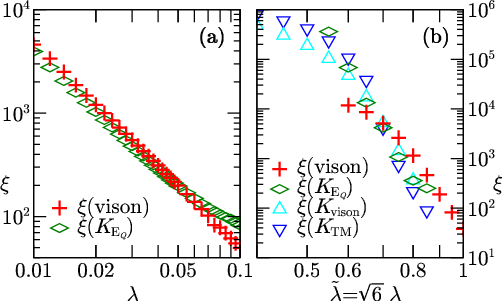}
\caption{Correlation length $\xi(\mathrm{vison})$ in the vison sector,
and fits $\xi(K)$ obtained from the sine-Gordon
model. \textbf{(a)} Dimer-solidomer interpolation. \textbf{(b)}~RVB model
with LR-singlets, using the $K$ of Fig.~\ref{fig:K}.
}
\label{fig:xi}
\end{figure}

From this discussion, we conclude that the LR-doped RVB model should be
gapped, but with a very large correlation length.  In order to verify
these conclusions, we have carried out high-precision 
simulations using boundary iMPS with bond dimension up to $\chi\approx 2500$  using the $\mathrm{SU}(2)$
symmetry and careful state-of-the-art
extrapolations~\cite{rams:epsilon-delta-extrapol}. We indeed find finite
correlation lengths for all values of $\lambda>0$ 
for which we obtain
reliable
extrapolations, down to $\tilde\lambda=0.6$,
where we
obtain a correlation length of $\xi\approx 1.2\times 10^4$,
see Fig.~\ref{fig:xi}; below that value, the
extrapolation becomes unreliable.  
In order to quantitatively compare with the field theory prediction, we first note 
that Eq.~\eqref{eq:sinegordon} is in fact the sine-Gordon model, for which
the
gap scaling $\xi = f(K) (\sqrt\omega\lambda)^{-u}$ with an explicit $f(K)$ is
analytically known~\cite{lukyanov:sine-gordon}.
To determine $\omega$, we can use that in the UV of
the field theory (i.e.\ below $\xi$, but above the lattice spacing), the
correlations scale as $\langle \cos(2\theta)(\vec x)\cos(2\theta)(\vec
y)\rangle=\tfrac12|x-y|^{-2/K}$~\cite{lukyanov:sine-gordon}. 
On the lattice, $\cos(2\theta)$ corresponds to ket-bra pairs of
$Q$ tensors; for the dimer-solidomer model, those have to sit on top of
each other, that is, $\mathbb E_Q \simeq \omega \cos(2\theta)$,
which allows to determine $\omega\equiv\omega(\lambda)$ from fitting the $\mathbb
E_Q$-correlator. This way, we obtain a prediction for $\xi$ without any free
parameters (in particular, a potential lattice spacing $a$ drops
out, as it then also appears in $\mathbb E_Q\simeq a^2\omega\cos(2\theta)$).
Fig.~\ref{fig:xi}a shows the prediction for the dimer-solidomer
interpolation, which demonstrates a remarkable agreement
of the measured data with the field theory prediction, in particular when
taking into account
that no free parameters were used. 
For the doped RVB, fitting $\omega$ is considerably more difficult, as the
ket-bra pairs of $Q$ tensors, which in the IR amount to $\cos(2\theta)$, 
no longer have to sit on top of each other, even though they remain bound
together. 
As a consequence, it is highly ambiguous how the pairs should
be binned in order to give $\omega\cos(2\theta)$ for any given unit volume.
An ad-hoc attempt would be to again use the $\mathbb E_Q$-correlator
to extract $\omega$, assuming strong binding of the pairs. However, this gives rise to correlations which are
about a factor $160$ too small, even though they exhibit the correct
scaling with $\tilde\lambda$, implying
that the resulting $\omega$ is too large. This suggests that including 
terms with nearby $Q$ leads to cancellations, which is plausible
given the antiferromagnetic nature of the RVB state. Unfortunately,
however, there is no unique way in which to group these terms, making a reliable
extraction of $\omega$ impossible.
Furthermore, the uncertainty of $\omega$ is amplified in $\xi$, since $\xi\propto
\omega^{-u/2}$, with $u\approx 6$ for the RVB.
In order to still be able to test the CFT prediction, we make the
assumption that
$\omega\equiv \bar\omega$ is
independent of $\tilde\lambda$. We can then fit this $\bar\omega$
by extracting $\omega(\tilde\lambda)$ from the measured correlation length
$\xi$, and taking $\bar\omega$ to be its average over the range
$\tilde\lambda =0.6,\dots,0.8$.  The resulting data is shown in
Fig.~\ref{fig:xi}b; in the light of the 
discussed uncertainty in extracting $\omega$, the overly simplistic
assumption of a constant $\bar\omega$, as well as other error
sources 
(non-relevant terms might renormalize interactions, higher orders in $\lambda$ are no
longer negligible), the agreement between the measured $\xi$ and the CFT
prediction is still surprisingly good.
Similar behavior is observed for $\lambda<0$, though generally we observe
a worse convergence.
Let us note that the dominant correlations in the doped RVB are vison
correlations (i.e., correlations obtained by placing a $Z$ string in the
PEPS), corresponding to the dominant solitons in the sine-Gordon model.
The correlations in the spinon sector, on the other hand, are 
short-ranged even at the $\mathrm{U}(1)$ point. In particular,
this implies that the
doping with long-range singlets induces a gapped topological spin liquid
phase with Toric Code order~\cite{iqbal:anyon-orderpars-tc-field},
in accordance with the findings of Ref.~\cite{chen:topo-z2-u1}.

Through this field theoretic treatment, we 
thus arrive at the following picture:
Perturbing a PEPS with $\mathrm{U}(1)$ entanglement symmetry in a way
which breaks the symmetry to $\mathbb Z_2$ immediately opens up a gap whenever
$K>\tfrac12$ in the effective field theory.
However, if $K$ is close to $\frac12$ and the perturbation is
sufficiently small, 
the correlation length $\xi\sim \lambda^u$ in the system is extremely large, as
$u=1/2(K-\tfrac12)+1$.
(At $K=\tfrac12$, the system is at a KT point where the correlation length
diverges superpolynomially.)
Thus, at scales
significantly below $\xi$, but above the lattice spacing (i.e., the UV of the
sine-Gordon field theory), the model essentially behaves like a critical
Lorentz-invariant theory. 
At these length scales, PEPS can therefore provide an accurate description
of critical systems without enforcing a virtual $\mathrm{U}(1)$
symmetry~\cite{wang:rvb-square-lattice,chen:topo-z2-u1}. 
In fact, the best
variational PEPS for critical systems can exhibit a rather short
correlation length which yet allows for reliable
extrapolations~\cite{rader:peps-ising-scaling,czarnik:finite-xi-scaling-iPEPS,corboz:finite-corrlength-scaling-iPEPS};
it is rather that the very large
correlation length we found makes the ansatz hard to converge, which 
suggests that breaking either lattice or $\mathrm{SU}(2)$ symmetries might
in fact result in more stable simulations.
Note that while guided by the RVB with longer-range singlets, these
findings in fact apply to arbitrary PEPS with $\mathrm{U}(1)$ entanglement
symmetries. In particular, this raises the question whether
$\mathrm{SU}(2)$ or $\mathrm{SU}(N)$ PEPS ansatzes which display chiral
features are truly critical (and chiral) away from special
$\mathrm{U}(1)$-invariant
points~\cite{poilblanc:kl-peps-1,poilblanc:kl-peps-2,hackenbroich:chiral-su2,chen:peps_su2_2_csl,chen:su3_peps_csl}. 

Finally, let us return to another one of the original motivations of this work,
namely to understand the possibility of emergent symmetries in PEPS.
As we have seen, perturbing the $\mathrm{U}(1)$ symmetry led to breakdown
of criticality, and thus it is safe to assume that the $\mathrm{U}(1)$
symmetry does not re-emerge under renormalization. But does this preclude
the possibility of emergent $\mathrm{U}(1)$ symmetries in 2D PEPS
altogether?  One way around would be to consider $\mathrm{U}(1)$-breaking
perturbations with higher magnetic flux $m$, while not increasing $K$; a
possibility for achieving that would be to consider doping of the RVB with
$N$-mers (trimers, tetramers, etc.).

\begin{figure}
\includegraphics{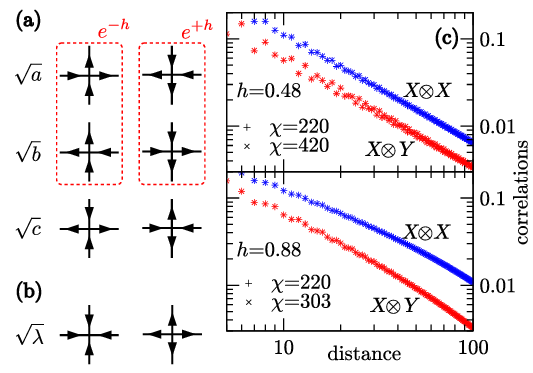}
\caption{PEPS with emergent $\mathrm{U}(1)$ symmetry. \textbf{(a)}
$6$-vertex model with magnetic field $h$; we choose $a=b=1$, $c=3$. The
corresponding PEPS (the superposition of all $6$-vertex configurations
with the indicated amplitude) has $K<\tfrac12$ for suitable $h$.
\textbf{(b)} Perturbation breaking the $\mathrm{U}(1)$ symmetry.
\textbf{(c)} Decay of correlations related by $\mathrm{U}(1)$ but not
$\mathbb Z_2$ symmetry. The identical decay scaling at 
$h=0.48$ ($K=0.435$) is consistent with an emergent $\mathrm{U}(1)$
symmetry in the correlations, while the different decay at $h=0.88$
($K=0.593$) rules it out. The data is converged in the corner transfer
matrix (CTM) bond dimension $\chi$. }
\label{fig:emergent}
\end{figure}

In the following, we present another approach and provide evidence
for an emergent $\mathrm{U}(1)$ symmetry. As we have observed, the scaling
dimension of the $\mathbb Z_2$ perturbation with $m=2$ is $1/K$, and thus
for $K<\tfrac12$ should become irrelevant. To construct a PEPS with
$K<\tfrac12$, we consider the $6$-vertex model with magnetic field $h$,
where each up-pointing (down-pointing) arrow acquires an additional
amplitude $e^{-h/2}$ ($e^{h/2}$), see Fig.~\ref{fig:emergent}a. The fixed
point of its transfer matrix is the ground state of the XXZ model with a
field, for which $K<\tfrac12$ for
suitable parameter
choices~\cite{baxter:book,verresen:stable-ll-and-emergent-u1} (in
particular, there is an finite range of values for $h$ where $K<1/2$, cf.\
Fig.~A2 of Ref.~\cite{verresen:stable-ll-and-emergent-u1},
and thus, the same will hold true for the corresponding
$6$-vertex model).
The corresponding PEPS is the superposition of all six-vertex configurations
with the corresponding weight, obtained by building tensors which carry
the arrow configurations both at the physical and the virtual degree of
freedom with the corresponding amplitude.  We have numerically studied the
effect of a perturbation 
which breaks $\mathrm{U}(1)$ to $\mathbb Z_2$
with amplitude $\sqrt\lambda$,
Fig.~\ref{fig:emergent}b. We choose $h=0.48$ and $h=0.88$, where we find 
$K=0.435$ and $K=0.593$, respectively.
The field theory predicts that in the former case 
we should have an emergent $\mathrm{U}(1)$ symmetry
at low energies and long distances.
To probe this, we measure the
correlation functions of two observables which are related by
$\mathrm{U}(1)$
symmetry at $\lambda = 0$. Specifically, we choose to put Paulis $X\otimes X$ and
$X\otimes Y$ on a ket/bra pair of bonds (the tensor product $A\otimes B$
denotes an operator $A$
on the ket bond and  another operator $B$ on the bra bond at a given
position); in the field
theory, these correspond to $\cos(\theta)$ and $\sin(\theta)$,
respectively. These are related by $\theta \to \theta+\pi/2$, which is no
longer a symmetry for $\lambda > 0$. Nevertheless, we find in
Fig.~\ref{fig:emergent}c that their correlation functions decay with the
same power law for $K<1/2$ ($h=0.48$), as opposed to $K>1/2$
($h=0.88$).  In the former case, this leaves the possibility to
construct linear combinations of (quasi-)local operators which yield
$\mathrm{U}(1)$-invariant correlation functions in the IR, while in the
latter case, this is impossible. Note that the emergent $\mathrm{U}(1)$
symmetry does \emph{not} require the two correlators to lie on top of each
other: The lattice operators can generate the field theory operators
$\cos\theta$ and $\sin\theta$ with different prefactors, which depend on
non-universal short-distance properties.
This observation is thus consistent with the predicted emergent
$\mathrm{U}(1)$ symmetry.
These results constitute the first evidence of emergent symmetries in
PEPS, which will be investigated further in future work.  In particular,
it remains to be seen whether such an emergent $\mathrm{U}(1)$ symmetry in
the correlators can be traced back to an emergent virtual $\mathrm{U}(1)$
symmetries of the PEPS tensors themselves, such as recently observed
for MPS e.g.\ at deconfined quantum critical
points~\cite{yang:emergent-syms}.

\emph{Acknowledgements.---}We acknowledge helpful discussions with
I.~Affleck,
P.~Fendley,
G.~Giudici,
G.~Ortiz,
F.~Pollmann, 
and
S.~Sondhi.
HD, JYC and NS acknowledge support from the European Union's
Horizon 2020 program through the ERC-StG WASCOSYS (No.~636201), and from the
DFG (German Research Foundation) under Germany's Excellence Strategy
(EXC2111-390814868).  
LV is supported by the Research Foundation Flanders.
RV is supported by the Harvard Quantum Initiative Postdoctoral
Fellowship in Science and Engineering, and a grant from the Simons
Foundation (\#376207, Ashvin Vishwanath).
The research of NS was funded in part by the Austrian Science Fund FWF
(Grant DOIs 
\href{https://doi.org/10.55776/P36305}{10.55776/P36305} and
\href{https://doi.org/10.55776/F71}{10.55776/F71}) 
and the European
Union’s Horizon 2020 research and innovation programme through Grant No.\
863476 (ERC-CoG SEQUAM).  Computations have been in part performed using
the Vienna Scientific Cluster (VSC).  For open access purposes, the
authors have applied a CC BY public copyright license to any author
accepted manuscript version arising from this submission.

\end{document}